\documentclass[aps,pre,reprint,groupedaddress]{revtex4-1}

\usepackage{amssymb,amsmath,bm,graphicx}
\usepackage{color}

\begin{document}

\title{Nonequilibrium Thermodynamics of the Soft Glassy Rheology Model}

\author{Ingo Fuereder, Patrick Ilg}
\affiliation{ETH Zurich, Department of Materials, Wolfgang-Pauli-Str.\ 10, CH-8093 Zurich}


\begin{abstract}
The Soft Glassy Rheology (SGR) model is a mesoscopic framework which proved to be very successful in describing flow and deformation of various amorphous materials phenomenologically (e.g. pastes, slurries, foams etc). In this paper, we cast SGR in a general, model independent framework for nonequilibrium thermodynamics called General Equation for the Nonequilibrium Reversible-Irreversible Coupling (GENERIC). This leads to a new formulation of SGR which clarifies how it can properly be coupled to hydrodynamic fields, resulting in a thermodynamically consistent, local, continuum version of SGR. Additionally, we find that compliance with thermodynamics imposes the existence of a modification to the stress tensor as predicted by SGR.
\end{abstract}


\maketitle


\section{Introduction}

Phenomenological models play an important role in understanding deformation and flow behavior of a large class of amorphous materials \cite{PhysRevLett.81.2934,PhysRevE.66.051501,PhysRevLett.103.036001,C1SM05229B,PhysRevE.78.016109,Pouliquen28122009}. One of these models is the so-called soft glassy rheology (SGR) model \cite{SGR}, which is frequently used to interpret experimental results on soft glassy materials including but not limited to pastes, slurries, colloidal suspensions, foams and polymeric dispersions  (e.g.\ \cite{Orsi201271,C2SM06889C,PhysRevLett.107.268302,C0SM00839G}). The SGR model has been investigated in some detail with respect to its predictions concerning rheological, diffusive and ageing behavior of soft glassy materials \cite{DiffSGR,PhysRevE.58.738,fielding:323}.

However, an aspect which has attracted attention only recently, is the issue of a proper thermodynamic interpretation of SGR. This problem was addressed in \cite{PhysRevE.85.031127} where the authors adapted a reasoning introduced by Bouchbinder and Langer in \cite{PhysRevE.80.031131,PhysRevE.80.031132,PhysRevE.80.031133}.
This analysis provided a new way of looking at an important parameter of the SGR model, namely the effective temperature $x$, and led to constraints on its time evolution.
However, the proof of the thermodynamic consistency of SGR as given in \cite{PhysRevE.85.031127} relies on specific assumptions concerning the entropy production and the form of the entropy itself. There, the authors note that it might be advantageous to address the question of thermodynamic consistency by casting SGR in a model-independent framework for nonequilibrium thermodynamics.

In this work we will follow this route by formulating SGR within a framework called General Equation for the Nonequilibrium Reversible-Irreversible Coupling (GENERIC) \cite{hco_book}. This has the following three benefits. Firstly, GENERIC allows us to prove the thermodynamic consistency of SGR based on more general assumptions compared to those made in \cite{PhysRevE.85.031127}. Secondly, we naturally obtain a closed set of time evolution equations for both the SGR degree of freedom and hydrodynamic fields allowing for a local, continuum description of SGR embedded in hydrodynamic flow in three dimensions as it is not present in the literature (although there is a tensorial version of SGR \cite{cates:193}, thermodynamical aspects and the coupling between SGR and hydrodynamics is not treated on this general level to the best of our knowledge). Thirdly, the structure of GENERIC allows us to identify a correction to the stress tensor as it is predicted by the standard SGR model.

We begin our discussion by briefly summarizing the SGR model and the GENERIC framework (Secs.\ II and III). In Sec.\ IV, we cast SGR in GENERIC form. Finally, several implications of this new formulation of SGR are discussed in Sec.\ V.

\section{The Soft Glassy Rheology Model}

The SGR model describes a glassy material as a collection of mesoscopic elements, containing several, cooperatively acting particles (e.g.\ a collection of cells in a foam). A single element is located in an energy landscape and trapped in a potential well of depth $E$. Due to thermal activation, one element might ``hop'' to another trap elsewhere in the energy landscape. On the microscopic level, this involves a local rearrangement of cooperative particles somewhere in the material. These rearrangements lead to stress redistributions in the material which might facilitate another rearrangement elsewhere. Many of these events are believed to sum up to an effective thermal noise level. This is accounted for by introducing an effective temperature $x$ which activates the ``hopping" process of the mesoscopic elements.
Rearrangements can additionally be facilitated by applying a macroscopic deformation to the sample where an element experiences a local strain $l$. It is assumed that every element behaves elastically supporting a local stress $kl$, where $k$ is an elastic constant.
If the stored elastic energy reaches the same order of magnitude as the trap depth, a yield event takes place (i.e.\ the system hops out of a trap). The attempt rate of escaping a trap is denoted by $\Gamma$ and is originally assumed to be of the following form: $\Gamma(E,l)=\Gamma_{0} \exp \left( -(E-kl^{2}/2)/x \right)$. Here, $\Gamma_{0}$ is a rate constant and we use units in which $k_{B}=1$ holds, such that temperatures are measured in energy units. After the occurrence of a yield event, the system rearranges and ends up in a new trap drawn from an a priori distribution $\rho(E,l)$ which models the presence of structural disorder. The SGR model was originally formulated for one dimension, where $\rho$ is assumed to be given by $\rho \sim \exp \left( -E/T_{g} \right) \delta(l)$.
Here $T_{g}$ is a glass transition temperature and the delta function sets the local strain variable to zero after a rearrangement has occurred. Finally, we denote the probability distribution function of $E$ and $l$ at time $t$ by $\psi$. The total stress in a soft glassy material as predicted by SGR is then given by the average of stress contributions from every element:
\begin{equation}
\label{stressvanilla}
     \sigma(t)=k \langle l \rangle_{\psi} = k \int l \psi(E,l) \mathrm{d} E \mathrm{d}l \ .
\end{equation}

This constitutive equation has to be supplemented by a time evolution equation for the probability distribution function $\psi$. This time evolution equation as stated by the SGR model contains a convective term, which accounts for the transportation of $\psi$ in $l$-space, and relaxation terms modeling the ``jump''-rate and the hop into a new trap. Thus, the overall time evolution reads as follows.

\begin{equation}
\label{tevvan}
 \partial_{t}  \psi = - \dot{\gamma} \partial_{l} \psi -\Gamma(E,l)\psi+Y\rho(E,l) \ ,
\end{equation}
where $\dot{\gamma}$ is the strain rate and $Y=\int{\Gamma(E,l) \psi \text{d}E\text{d}l }$ the overall yielding rate for a given $\psi$.
With the previously mentioned ansatz for $\Gamma$ and $\rho$, the equilibrium solution for this one-dimensional version of SGR becomes

\begin{equation}
\label{equsol}
  \psi_{0}^{1\text{D}} \sim  \exp \left( -E/T_{g}+E/x \right) \delta(l)  \ ,
\end{equation}
which ceases to be normalizable for temperatures $x < T_{g}$; an equilibrium state does not exist anymore in this temperature regime and the system shows various aging phenomena \cite{fielding:323}.  Together with proper initial conditions, equations,  (\ref{stressvanilla}) and (\ref{tevvan}) fully determine the stress in a soft glassy material. It is obvious, that equation ($\ref{stressvanilla}$) and the choice for the prior distribution $\rho$ cannot trivially be translated in a three dimensional model. As it will be discussed later, a slightly different choice for $\rho$ is necessary in three dimensions, but the main features of the original SGR model remain the same in more general versions of it.

\section{GENERIC}

Since GENERIC is a model-independent framework for nonequilibrium thermodynamics, it proved to be a very powerful tool for investigating the compliance of dynamic equations of various models with thermodynamics. GENERIC is a formulation of nonequilibrium thermodynamics which divides the dynamics of a closed system into two parts \cite{PhysRevE.56.6620,PhysRevE.56.6633}. Its first part is the reversible contribution describing the purely mechanistic motion whereas the second part is accounting for irreversible dynamics. The framework implies a description in terms of a set of carefully chosen slowly-evolving variables assuming a clear time-scale gap between these variables and fast (irrelevant) degrees of freedom. For simple fluids, densities of conserved quantities (i.e.\ mass density, momentum density and energy density) are appropriate variables to consider. For complex fluids we introduce structural variables in addition to the hydrodynamic variables. They need to be chosen such that they contain enough information of the system's state to determine stresses without any memory effects. Hence, if $\bm{x}$ denotes a set of variables which appropriately describe the system, their time evolution can be cast into the following form

\begin{equation}
\label{GEN1}
      \frac{d\bm{x}}{dt}=L(\bm{x}) \cdot \frac{\delta E(\bm{x})}{\delta \bm{x}} + M(\bm{x}) \cdot \frac{\delta S(\bm{x})}{\delta \bm{x}} \ .
\end{equation}

Here, $\delta/\delta \bm{x}$ denotes the functional derivative of the energy and entropy functionals, meaning that energy gradients drive the reversible particle motion, whereas entropy gradients generate irreversibility. The linear operators $L(\bm{x})$ (Poisson matrix) and $M(\bm{x})$ (friction matrix) represent the geometric (Poisson) structure underlying the reversible motion and the dissipative material properties, respectively. Their action on the energy and entropy gradient involves an additional integration over the system's volume wherever fields are involved.
Associated with the Poisson matrix and two observables, $A$ and $B$  (i.e.\  real valued, sufficiently regular functionals of the set of variables $\bm{x}$), is a Poisson bracket,

\begin{equation}
\label{GEN2}
      \{A,B\}:= \frac{\delta A(\bm{x})}{\delta \bm{x}} \cdot L(\bm{x}) \cdot  \frac{\delta B(\bm{x})}{\delta \bm{x}} \ ,
\end{equation}
satisfying an antisymmetry condition $\{A,B\}= -\{B,A\}$. Furthermore, using a third observable $C$, the bracket obeys the Leibniz rule, $\{AB,C\}= A\{B,C\} +B\{A,C\}$, and the Jacobi Identity, $\{A,\{B,C\}\} +  \{B,\{C,A\}\} +  \{C,\{A,B\}\} = 0$.

These conditions pose severe restrictions on the admissible form of $L(\bm{x})$ and therefore on the convective behavior of $\bm{x}$. Similarly, a dissipative bracket is defined for the friction matrix, $M(\bm{x})$:

\begin{equation}
\label{GEN6}
    [A,B]:= \frac{\delta A(\bm{x})}{\delta \bm{x}} \cdot M(\bm{x}) \cdot  \frac{\delta B(\bm{x})}{\delta \bm{x}} \ ,
\end{equation}
being symmetric, $[A,B]=[B,A]$, and positive, $[A,A] \geq 0$.

The Poisson bracket and dissipative bracket allow us to write the time evolution of an arbitrary observable $A$  in a compact form,

\begin{equation}
\label{GEN9}
      \frac{dA}{dt}=  \{A,E\}+  [A,S] \ .
\end{equation}

This equation is supplemented by the degeneracy requirements

\begin{equation}
\label{GEN10}
      L(\bm{x}) \cdot  \frac{\delta S(\bm{x})}{\delta \bm{x}}=0
\end{equation}
and

\begin{equation}
\label{GEN11}
      M(\bm{x}) \cdot  \frac{\delta E(\bm{x})}{\delta \bm{x}}=0 \ .
\end{equation}
These degeneracy conditions together with the symmetries of the brackets guarantee that the energy is conserved,

\begin{equation}
\label{GEN12}
      \frac{dE}{dt}=  \{E,E\}+  [E,S] = 0  \,
\end{equation}
and that entropy is a non-decreasing function of time,

\begin{equation}
\label{GEN13}
      \frac{dS}{dt}= [S,S]  \ge 0  \ .
\end{equation}
Thus, we note that GENERIC is by construction sufficient for equations (\ref{GEN12}) and (\ref{GEN13}) but obviously not necessesary for them. However, the generality of the assumptions discussed in this section ensure its applicability to a wide class of models.
A more detailed discussion of the GENERIC structure and example applications can be found in \cite{hco_book}.
We now implement a GENERIC formulation of SGR by choosing an appropriate set of variables and constructing the GENERIC building blocks $E$, $S$, $L$ and $M$.

\section{GENERIC formulation of SGR}

An appropriate choice of the variables $\bm{x}$ is \textit{the} crucial first step and a prerequisite for formulating a constitutive model within the GENERIC framework. Since we are interested in a formulation of SGR which allows for a proper treatment of hydrodynamics, it is natural to include the conserved quantities mass density $\rho_{\text{m}}$, momentum density $\bm{m}$ and energy density in the set of variables. Following the reasoning presented in \cite{PhysRevE.85.031127} and \cite{PhysRevE.80.031131,PhysRevE.80.031132,PhysRevE.80.031133}, we conceptually divide the degrees of freedom in a soft glassy material in two coupled subsystems: A configurational subsystem which describes in which one of the local energy minima of the potential energy landscape (inherent structures) the system currently is in and a kinetic-vibrational subsystem which accounts for the energy contribution arising from the motion around these minima. Additionally, the authors in \cite{PhysRevE.85.031127} and \cite{PhysRevE.80.031131,PhysRevE.80.031132,PhysRevE.80.031133} consider a third subsystem being a thermal reservoir and setting the thermodynamic temperature of the system. For the sake of simplicity, we do not distinguish between the reservoir and the fast kinetic-vibrational degrees of freedom tacitly assuming a strong coupling between these subsystems. Therefore, we account for the configurational and fast contributions to the energy by adding an energy density for both the configurational subsystem ($\epsilon_{\text{c}}$) and the kinetic-vibrational subsystem ($\epsilon$) in our list of variables. Finally, the last state variable is the SGR probability distribution function $\psi$ itself, which accounts for the barrier height $E$ felt by the mesoscopic SGR-elements and for their local strain $\bm{l}$.
Thus, our total set of variables for the SGR model is given by

\begin{equation}
\label{GSGR1}
       \bm{x}=\left\{\rho_{\text{m}}(\bm{r}),\bm{m}(\bm{r}),\epsilon(\bm{r}),\epsilon_{\text{c}}(\bm{r}),\psi(E,\bm{l},\bm{r})  \right\} \ .
\end{equation}

We allow all quantities to depend on the position in the system $\bm{r}$ though the length scale of interest has to be taken much larger than the dimension of one SGR element. On this coarser scale every hydrodynamic volume element contains enough SGR elements to have a meaningful local distribution of yield energies $E$ and strains $\bm{l}$.
Here, we describe the strain of a SGR element with a vector $\bm{l}$ implicitly considering those elements as dumbbell-like objects as it was done in previous tensorial formulations of SGR \cite{cates:193}. All quantities are local in nature and we will suppress the position argument in our notation from now on for the sake of simplicity.

\subsection{Total Energy and Total Entropy}

With our previous considerations at hand, it is straightforward to formulate a functional for the total energy $E[\bm{x}]$,

\begin{align}
\label{GSGR2}
 E[\bm{x}]= & \int{ \left( \frac{\bm{m}^2}{2\rho_{\text{m}}} + \epsilon +\epsilon_{\text{c}} \right)} \text{d}^3\bm{r} \nonumber \\
& + \int{\phi^{E}(E,\bm{l})\psi(E,\bm{l})} \text{d}E \text{d}^3\bm{l}  \text{d}^3\bm{r} \ .
\end{align}

\begin{figure}
 \includegraphics[scale=0.7]{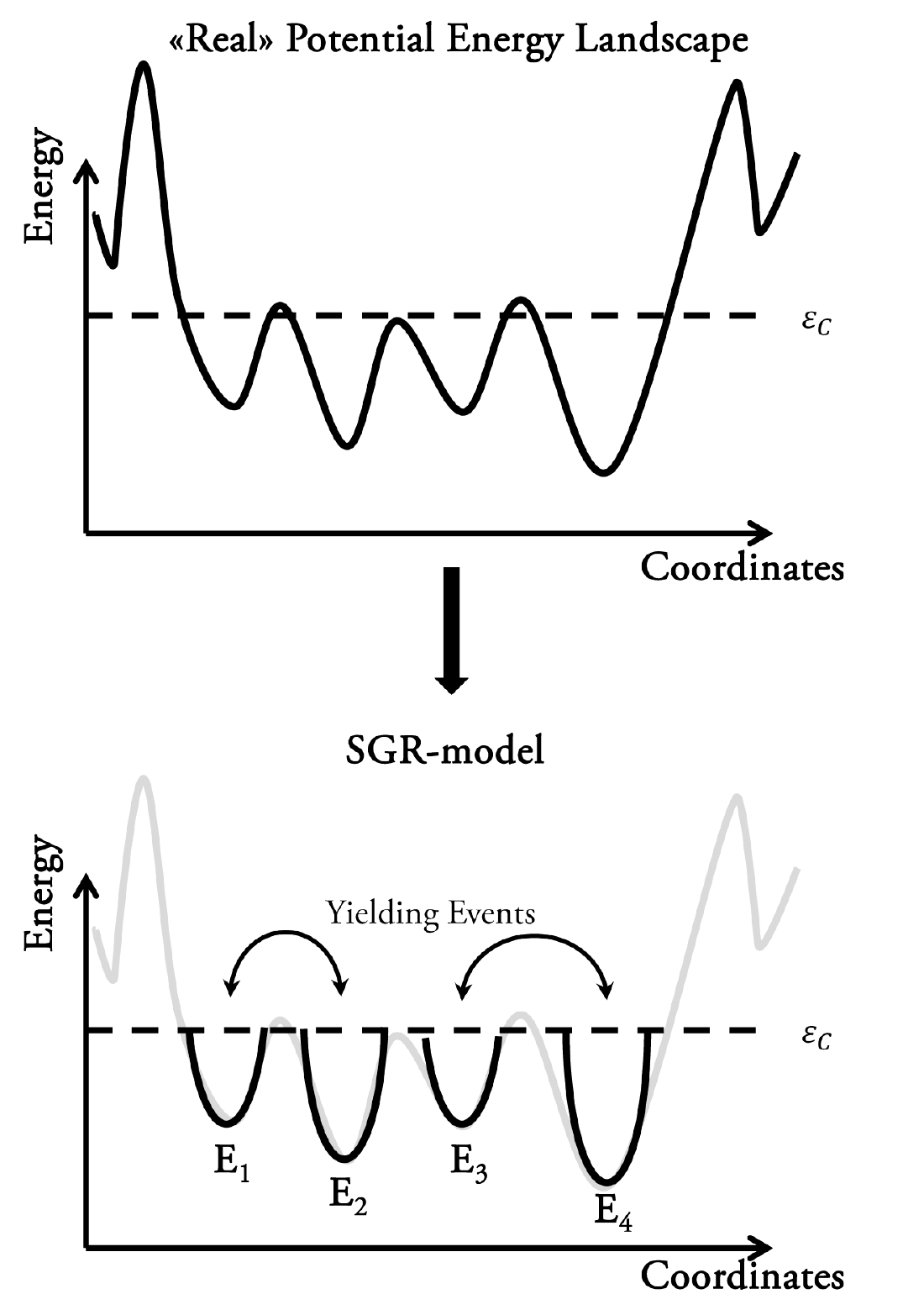}%
 \caption{\label{fig}The variable $\epsilon_{\text{c}}$ sets a level in the potential energy landscape. Within the SGR-model wells below this level are approximated by harmonic traps with a trap depth $E_{i}$ drawn from the distribution $\rho(E,\bm{l})$. Work performed on the system by deforming the material can be stored by the mesoscopic SGR-elements. If the stored elastic energy becomes comparable to the energy depth of a well, a yielding event takes place.}
 \end{figure}

The first two terms are the kinetic energy accounting for the flow of the fluid and the internal energy density of the fast subsystem $\epsilon$. The energy density of the configurational subsystem $\epsilon_{\text{c}}$ has to be regarded as a level in the ``true" potential energy landscape which defines the region of this landscape being accessible for the system. The wells below this level are approximated by harmonic traps according to the SGR model (see Fig.$\ref{fig}$).
The last term in the total energy accounts for these traps, where we introduced a potential $\phi^{E}(E,\bm{l})$, describing the energy gain for every mesoscopic element residing in the bottom of a well. This energy gain is given by the trap depth $-E$. However, every SGR-element can elastically be strained, which  increases the effective trap depth by the stored elastic energy $k\bm{l}^2/2$. These considerations lead us to the assumption $\phi^E(E,\bm{l})=k \bm{l}^2/2-E$. We note, that a conceptually similar model was developed in \cite{hcogrmelahutter}. There, the authors describe plastic deformation of single crystalline materials by considering a periodic arrangement of identical energy wells. The dynamic variable of this model is given by the distribution function of the strain between two layers of the material and the plastic strain rate emerges as the average hopping rate between energy wells.

In a next step, we make an ansatz for the total entropy,

\begin{align}
\label{GSGR3}
S[\bm{x}]= & \int{ \bigg[ s \left( \rho ,\epsilon \right) + s_{\text{c}}(\epsilon_{\text{c}}) \bigg] \text{d}^3\bm{r}} \nonumber \\
& -\int{\psi(E,\bm{l})\ln\frac{\psi(E,\bm{l})}{R (E,\bm{l})}\text{d}E\text{d}^3\bm{l}\text{d}^3\bm{r}} \ ,
\end{align}
where we separate entropy contributions from the reservoir ($s$) and  from the configurational subsystem ($s_{\text{c}}$) implicitly making a local equilibrium assumption for both the fast and the slow subsystem. Additionally, we made a conventional ansatz for the form of the $\psi$-dependence of the entropy. At this point, we keep a general function $R(E,\bm{l})$ and postpone a specific choice for the reference distribution function.
The derivatives of total energy and total entropy with respect to the state variables are given by

\begin{equation}
\label{GSGR4}
   \frac{\delta E}{\delta \bm{x}} = \left(-\frac{\bm{v}^2}{2},\bm{v},1,1,\phi^{E} \right)
\end{equation}
and
\begin{equation}
\label{GSGR5}
   \frac{\delta S}{\delta \bm{x}} = \left(-\frac{\mu}{T},0,\frac{1}{T},\frac{1}{\chi},-\ln\frac{\psi}{R} \right) \ .
\end{equation}
Here, we introduce the local fluid velocity $\bm{v}=\bm{m}/\rho_{\text{m}}$ and the chemical potential $\mu$.
Furthermore, we define the standard thermodynamic temperature $T$ and a configurational temperature $\chi$ via $1/T= \partial s/ \partial \epsilon$ and $1/\chi= \partial s_{c}/ \partial \epsilon_{c}$.
\subsection{The Poisson Matrix $L$}

Having formulated expressions for energy and entropy, we proceed to the construction of the Poisson matrix.
The procedure of properly implementing the Poisson matrix for standard hydrodynamics was worked out in one of the original publications on GENERIC \cite{PhysRevE.56.6633}. We note that the specific form of the entries in the matrix is fully determined by the tensorial character of the state variables \cite{hco_book}. Therefore, the construction of the Poisson matrix is straightforward and reads as follows:

\begin{widetext}
\begin{equation}
\label{GSGR6}
 L=-\begin{pmatrix} 0 & \frac{\partial}{\partial \bm{r}}\rho_{\text{m}} & 0 & 0 & 0 \\
  \rho_{\text{m}}\frac{\partial}{\partial \bm{r}} & \left( \frac{\partial}{\partial \bm{r}} \bm{m}+\bm{m}\frac{\partial}{\partial \bm{r}} \right)^T & \epsilon \frac{\partial}{\partial   \bm{r}}+\frac{\partial}{\partial \bm{r}} \cdot \bm{\Pi}^{S} & \epsilon_{\text{c}} \frac{\partial}{\partial \bm{r}}+\frac{\partial}{\partial \bm{r}} \cdot \bm{\Pi}^{S}_{\text{c}}  &   \psi\frac{\partial}{\partial\bm{r}}-\frac{\partial}{\partial\bm{r}}\cdot \psi\bm{l}\frac{\partial}{\partial\bm{l}} \\
   0 & \frac{\partial}{\partial\bm{r}} \epsilon + \bm{\Pi}^S \cdot \frac{\partial}{\partial \bm{r}}   & 0 & 0 & 0 \\
   0 & \frac{\partial}{\partial\bm{r}} \epsilon_{\text{c}} + \bm{\Pi}^{S}_{\text{c}} \cdot \frac{\partial}{\partial \bm{r}}   & 0 & 0 & 0 \\
   0 &  \frac{\partial}{\partial\bm{r}} \psi+\frac{\partial}{\partial\bm{l}} \psi\bm{l} \cdot \frac{\partial}{\partial\bm{r}} & 0 & 0 & 0 \end{pmatrix} \ ,
\end{equation}
\end{widetext}
where the differential operators act on all functions to their right. The entry in the last row guarantees a proper convection of the distribution function $\psi$ itself and of the strain vector $\bm{l}$. We have allowed for entropic pressure tensor contributions  ($\bm{\Pi}^{S}$ and $\bm{\Pi}^{S}_{\text{c}}$) of reservoir and the configurational subsystem. Their form is determined by the degeneracy condition ($\ref{GEN10}$). It is only the multiplication of the second row with the entropy gradient which does not trivially satisfy the degeneracy condition but gives the condition

\begin{widetext}
\begin{equation}
\label{GSGR7}
    \rho_{\text{m}} \frac{\partial }{\partial \bm{r}} \frac{\mu}{T} -\epsilon \frac{\partial }{\partial \bm{r}} \frac{1}{T} - \frac{\partial }{\partial \bm{r}} \cdot \frac{\bm{\Pi}^{S}}{T} - \frac{\partial }{\partial \bm{r}} \cdot \frac{\bm{\Pi}^{S}_{c}}{\chi} +\int{\left( \psi \frac{\partial }{\partial \bm{r}} \ln \frac{\psi}{R}-\frac{\partial }{\partial \bm{r}} \cdot \psi \bm{l}\frac{\partial }{\partial \bm{l}} \ln \frac{\psi}{R} \right) } \text{d}E\text{d}^3\bm{l}  = 0 \ .
\end{equation}
\end{widetext}

Using the expression for the hydrostatic pressure $p$
\begin{equation}
\label{GSGR8}
    p=T\left( s-\rho_{\text{m}}\frac{\partial s}{\partial \rho_{\text{m}}}-\epsilon\frac{\partial s}{\partial \epsilon}\right) \ ,
\end{equation}
identifying the hydrodynamic pressure tensor with the pressure contribution arising from the standard background fluid $\bm{\Pi}^{S}$ (i.e.: $\bm{\Pi}^{S}=p\bm{1}$), assuming that $\ln R$ does not depend on position and performing an integration by parts, we finally find the following configurational contribution to the stress.

\begin{equation}
\label{GSGR9}
   \frac{\bm{\Pi}^{S}_{\text{c}}}{\chi} = \int{\psi(E,\bm{l})\left[2\bm{1}+\bm{l} \frac{\partial \ln R(E,\bm{l})}{\partial \bm{l}} \right] \mathrm{d}E }\mathrm{d}^3\bm{l} \ .
\end{equation}

The first term matches the hydrodynamic pressure $p$ and is not of interest for standard rheological measurements, whereas the second part is an additional entropic stress tensor contribution which has not been taken into account by previous versions of the SGR model. Its specific form depends on the reference distribution. We will discuss a natural choice for $R$ and its implications in the next section.

 \subsection{The Friction Matrix $M$}

We now turn to the friction matrix describing the dissipative processes present in our system. As we constructed our model, we expect the following three contributions to the $M$-matrix: Firstly, we assume that the reservoir acts like a standard Newtonian fluid (i.e.\ it contributes to the rheological response with a viscosity $\eta$ and a dilatational viscosity $\hat{\kappa}$). Additionally, we model its thermal behavior satisfying Fourier's law of heat conduction. A proper $M$-matrix for this hydrodynamic contribution ($M_{\text{HD}}$) was already given in \cite{PhysRevE.56.6633}. The second, configurational part ($M_{\text{C}}$) is the dissipative process of yield events as modeled by the non-convective part of the SGR time evolution equation for the probability density $\psi$. As we will discuss later, the SGR model can be viewed as a time continuous Markov process for which the appropriate form of the $M$-matrix is known as well \cite{Schieber2003179}. Finally, the last contribution ($M_{\text{HF}}$) can trivially be formulated and models the coupling between reservoir and configurational degrees of freedom allowing for an energy exchange/heat flow between those subsystems. This means that the form of the $M$-matrix for all involved dissipative contributions to our SGR formulation is well known. These contributions add up to the following total matrix:

\begin{widetext}
\begin{multline}
\label{GSGR10}
  M=\underbrace{\begin{pmatrix} 0 & 0 & 0 & 0 & 0\\ 0 &  -\frac{\partial}{\partial \bm{r}} \eta T \frac{\partial}{\partial \bm{r}} + \bm{1} \frac{\partial}{\partial \bm{r}} \cdot \eta T \frac{\partial}{\partial \bm{r}}-\frac{\partial}{\partial \bm{r}} \hat{\kappa} T \frac{\partial}{\partial \bm{r}} & \frac{\partial}{\partial \bm{r}} \cdot \eta T \dot{\bm{\gamma}} +\frac{\partial}{\partial \bm{r}}\frac{\hat{\kappa}T}{2}\mathrm{Tr}(\dot{\bm{\gamma}}) & 0 & 0 \\ 0 & -\eta T \dot{\bm{\gamma}}\cdot\frac{\partial}{\partial \bm{r}}-\frac{\hat{\kappa}T}{2} \mathrm{Tr}(\dot{\bm{\gamma}})\frac{\partial}{\partial \bm{r}} & \frac{\eta T}{2}\dot{\bm{\gamma}}:\dot{\bm{\gamma}}+\frac{\hat{\kappa}T}{4}\left(\mathrm{Tr}(\dot{\bm{\gamma}})\right)^{2}-\frac{\partial}{\partial \bm{r}} \cdot \bm{\lambda} \cdot T^{2}\frac{\partial}{\partial \bm{r}} & 0 & 0 \\ 0 & 0 & 0 & 0 & 0 \\ 0 & 0 & 0 & 0 & 0 \end{pmatrix}}_{M_{\text{HD}}} + \\
\underbrace{\begin{pmatrix} 0 & 0 & 0 & 0 & 0 \\ 0 & 0 & 0 & 0 & 0 \\ 0 & 0 & \alpha & -\alpha & 0 \\ 0 & 0 & -\alpha & \alpha & 0 \\ 0 & 0 & 0 & 0 & 0 \end{pmatrix}}_{M_{\text{HF}}} +
\underbrace{\begin{pmatrix} 0 & 0 & 0 & 0 & 0 \\ 0 & 0 & 0 & 0 & 0 \\ 0 & 0 & 0 & 0 & 0 \\ 0 & 0 & 0 & \phi^{E} M_{55} \phi^{E} & -M_{55} \phi^{E} \\ 0 & 0 & 0 & -M_{55} \phi^{E} & M_{55} \end{pmatrix}}_{M_{\text{C}}}
\end{multline}
\end{widetext}

Here, we denote a double contraction by a colon (e.g., $\dot{\bm{a}}:\dot{\bm{b}}=\sum\limits_{i,j}\dot{a}_{ij}\dot{b}_{ji}$). We used the rate of strain tensor $\dot{\bm{\gamma}}$ and allowed for a tensorial heat conduction tensor $\bm{\lambda}$ which can model anisotropic heat flow in the reservoir. The (positive) function $\alpha$ describes the energy exchange between the reservoir and the configurational subsystem. The operator $M_{55}$ contains all the information of the yielding processes. $M_{\text{C}}$ was constructed in the following way. First, an appropriate operator $M_{55}$  is chosen to describe the dissipative part of the SGR model. Then, the degeneracy condition (\ref{GEN11}) determines the entry $M_{54}$ which automatically is equal to $M_{45}$ due to the symmetry of $M$. The entry $M_{44}$ is again fixed by the degeneracy requirement. The other parts of the friction matrix are constructed analogously which guarantees, that the proposed $M$-matrix satisfies the degeneracy condition (\ref{GEN11}).  Our last step is formulating an expression for the operator $M_{55}$.
We note that the non-convective part of the SGR time evolution equation (\ref{tevvan}), $\dot{\psi}_{\text{nc}}$, can be viewed as a time continuous Markov process with the transition rate $w(E \rightarrow E',\bm{l} \rightarrow \bm{l}')=\Gamma(E,\bm{l})=\Gamma_{0} \exp \left[ -(E-k\bm{l}^{2}/2)/x \right]$ and the density of states $\rho(E',\bm{l}')$,

\begin{align}
\label{GSGR11}
 \dot{\psi}_{\text{nc}}(E,\bm{l}) = & -\int{ w(E \rightarrow E',\bm{l} \rightarrow \bm{l}') \rho(E',\bm{l}') \psi(E,\bm{l}) \text{d}E'\text{d}^3\bm{l}' } \nonumber  \\
& + \int{ w(E' \rightarrow E,\bm{l}' \rightarrow \bm{l}) \rho(E,\bm{l}) \psi(E',\bm{l}') \text{d}E'\text{d}^3\bm{l}' } \ .
\end{align}

It can easily be verified, that inserting the definition of the rate $w$ in ($\ref{GSGR11}$) indeed results in the nonconvective time evolution of the SGR distribution function.
Any entropy driven master equation, describing a Markovian jump process satisfying detailed balance
\begin{equation}
\label{GSGR12}
   w(E \rightarrow E',\bm{l} \rightarrow \bm{l}') \rho' \psi_{0}= w(E' \rightarrow E,\bm{l}' \rightarrow \bm{l}) \rho \psi_{0}'
\end{equation}
can be cast into GENERIC form \cite{Schieber2003179}. The corresponding operator $M_{55}$ is given by
\begin{multline}
\label{GSGR13}
  M_{55}=\int \Gamma(E,\bm{l}) \Gamma(E'',\bm{l}'')       \frac{\psi''\psi_{0}-\psi\psi_{0}''}{\ln\left( \frac{\psi''\psi_{0}}{\psi_{0}''\psi} \right)}  \times \\   \Big[\delta(E-E',\bm{l}-\bm{l'})-\delta(E''-E',\bm{l}''-\bm{l'}) \Big] \mathrm{d}E''\mathrm{d}^3\bm{l}'' \ ,
\end{multline}
where we used the abbreviation $\psi_{0}=Y\rho/\Gamma$, which is the equilibrium distribution function for $x>T_{\text{g}}$ . In equations ($\ref{GSGR12}$) and ($\ref{GSGR13}$) we did not explicitly include the variable dependency of the distribution function but used $\psi=\psi(E,\bm{l},\bm{r})$ and $\psi''=\psi(E'',\bm{l}'',\bm{r})$ as a short notation.
The operator $M_{55}$ acts on functions which depend on the variables $\left\{ \bm{r},E',\bm{l}' \right\}$ or $\left\{\bm{r},E,\bm{l}\right\}$ . Note that multiplication with $M_{55}$ involves an additional integration step over $E$ and $\bm{l}$ or $E'$ and $\bm{l}'$, respectively.

The construction of $M_{55}$ is detailed in the appendix. As discussed in Sec.\ III, we have to show that the proposed $M$-matrix in total is non-negative and symmetric in order to obtain a thermodynamically valid model. A discussion of these properties of the $M$-matrix can also be found in the appendix. This completes our construction of the GENERIC building blocks and we proceed with the discussion of the obtained time evolution equations. \\

\section{Time Evolution Equations}

In this section we combine all previously discussed GENERIC building blocks to obtain a thermodynamically consistent set of equations for the state variables $\bm{x}$.
To obtain the time evolution for $\psi$ we proceed as follows. First, we specify the function $\phi^{E}(E,\bm{l})$. As it was argued previously and following the reasoning given in \cite{PhysRevE.85.031127}, we expect this function to be nothing else than $k\bm{l}^{2}/2-E$, i.e.\ the effective trap depth felt by a strained SGR element. Additionally, we make a specific choice for the reference distribution $R$. A natural assumption is using the a priori distribution; $R(E,\bm{l})=\rho(E,\bm{l})$.
The irreversible part of the time evolution of $\psi$ is given by
\begin{multline}
\label{GSGR13a}
   M_{55} \left( -\frac{\phi^{E}(E',\bm{l}')}{\chi}+ \frac{\delta S}{\delta \psi'} \right) = \\  - M_{55}\left( \ln\frac{\psi(E',\bm{l}')}{R(E',\bm{l}') \exp(-\phi^{E}(E',\bm{l}')/ \chi)} \right) \ .
\end{multline}
If we want to obtain SGR-like equations in our GENERIC formulation, we have to postulate, that the denominator of the expression in the logarithm is proportional to the equilibrium distribution function, i.e.\  $R(E,\bm{l}) \exp(-\phi^{E}/\chi) \sim \psi_{0}$.  With our choice of $R$ , this proportionality  is true only if $\chi=x$, i.e.\ we assume that the configurational temperature coincides with the effective temperature. This observation is in agreement with conclusions drawn in \cite{PhysRevE.85.031127}. In total, this leads to
\begin{multline}
\label{GSGR20}
-M_{55} \ln \frac{\psi'}{\psi_{0}'} =  -\int \int  \Gamma(E,\bm{l}) \Gamma(E',\bm{l}')    \frac{\psi''\psi_{0}-\psi\psi_{0}''}{\ln\left( \frac{\psi''\psi_{0}}{\psi_{0}''\psi} \right)}    \times \\ \Big[\delta(E-E',\bm{l}-\bm{l'})-\delta(E''-E',\bm{l}''-\bm{l'}) \Big]\ln \frac{\psi'}{\psi_{0}'} \mathrm{d}E''\mathrm{d}^3\bm{l}''  \mathrm{d}E'\mathrm{d}^3\bm{l}' \\ =  \int  \Gamma(E,\bm{l}) \Gamma(E',\bm{l}')  \left(  \psi''\psi_{0}-\psi\psi_{0}'' \right) \mathrm{d}E''\mathrm{d}^3\bm{l}''   \ ,
\end{multline}
Inserting the definition for $\psi_{0}$ and for the rates $w$ we find that this is equal to the right side of equation ($\ref{GSGR11}$) and we obtain the non-convective time evolution for $\psi$.
The total set of time evolution equations reads as follows:
\begin{widetext}
\begin{align}
\label{GSGR21}
\dot{\rho}_{\text{m}}& =  - \frac{\partial}{\partial \bm{r}} \cdot \left( \rho_{\text{m}}\bm{v} \right) \\
\label{mtee}
\dot{\bm{m}}& = -\frac{\partial}{\partial \bm{r}} \cdot (\bm{m} \bm{v}) +  \frac{\partial}{\partial \bm{r}} \cdot \bm{\tau} \\
\dot{\epsilon}& =-\frac{\partial}{\partial \bm{r}} \cdot (\epsilon \bm{v})-\bm{\Pi}^{S}:\frac{\partial}{\partial \bm{r}}\bm{v}+\frac{\eta}{2}\dot{\bm{\gamma}}:\dot{\bm{\gamma}}+\frac{\hat{\kappa}}{4}\left[\mathrm{Tr}(\dot{\bm{\gamma}})\right]^{2}-\frac{\partial}{\partial \bm{r}} \cdot \bm{\lambda} \cdot T^{2}\frac{\partial}{\partial \bm{r}}\frac{1}{T} + \alpha \left(\frac{1}{T}-\frac{1}{\chi} \right) \\
\dot{\epsilon}_{c}& =-\frac{\partial}{\partial \bm{r}} \cdot (\epsilon_{\text{c}} \bm{v})-\bm{\Pi}^{S}_{\text{c}}:\frac{\partial}{\partial \bm{r}}\bm{v}+ \alpha \left(\frac{1}{\chi} -\frac{1}{T}\right)-\int{\phi^{E}(E,\bm{l}) \Big[ -\Gamma(E,\bm{l})\psi(E,\bm{l})+Y \rho(E,\bm{l}) \Big]  \text{d}E\text{d}^3\bm{l}} \\
\label{psitee}
\dot{\psi}&=-\frac{\partial}{\partial \bm{r}} \cdot \Big[\psi(E,\bm{l}) \bm{v}\Big]-\frac{\partial}{\partial \bm{l}} \cdot \left(\frac{\partial}{\partial \bm{r}}\bm{v}\right)\cdot \bm{l}\psi(E,\bm{l})-\Gamma(E,\bm{l})\psi(E,\bm{l})+Y \rho(E,\bm{l})
\end{align}
\end{widetext}
In equation ($\ref{mtee}$) we have introduced the total stress tensor,

\begin{align}
\label{GSGR21a}
\bm{\tau} = & -\bm{\Pi}^{S}- \bm{\Pi}^{S}_{\text{c}}+ \int{\left(\psi(E,\bm{l})\bm{l}\frac{\partial}{\partial\bm{l}}\phi^{E}(E,\bm{l})\right)}\mathrm{d}E\mathrm{d}^3\bm{l} \nonumber \\  & +  \eta \dot{\bm{\gamma}} +  \frac{\hat{\kappa}}{2}\mathrm{Tr}(\dot{\bm{\gamma}})  \bm{1} \nonumber \\ = & - \Big[p+2\chi - \frac{\hat{\kappa}}{2}\mathrm{Tr}(\dot{\bm{\gamma}}) \Big] \bm{1} +  \eta \dot{\bm{\gamma}} +k \int{\psi(E,\bm{l})\bm{l}\bm{l} }\mathrm{d}E\mathrm{d}^3\bm{l} \nonumber \\  & - \chi \int{\psi(E,\bm{l})\left[ \bm{l} \frac{\partial \ln R(E,\bm{l})}{\partial \bm{l}} \right] \mathrm{d}E }\mathrm{d}^3\bm{l}     \ .
\end{align}

We briefly discuss the physical meaning of the terms appearing in the equations of motion for the energy densities ($\epsilon, \epsilon_{\text{c}}$):
The first part on the right hand side of the equation for $\dot{\epsilon}$ is simply convective transport. The second term includes the hydrostatic pressure of the reservoir. The terms containing the viscosity constants describe viscous heating and the term containing the heat conduction tensor $\bm{\lambda}$ allows for heat transport in the fast subsystem. Finally, the last term describes heat transfer between fast and slow subsystem.
In the equation for $\dot{\epsilon}_{c}$, the heat transfer term between slow and fast subsystem appears again but with opposite sign accounting for heat flux in the reverse direction. The last term is the energy which is dissipated due to yield events.
Equation  ($\ref{psitee}$) perfectly coincides with the time evolution as stated by the dumbbell-like tensorial version of the SGR model \cite{cates:193} besides the presence of an additional term in our model which accounts for spatial convection.

\section{Conclusion and Discussion}

In this paper we have successfully formulated the SGR model within the GENERIC framework from which we draw the following conclusions: The SGR model is proven to be thermodynamically consistent, i.e.\ an isolated system of a soft glassy material in which the SGR equation of motion is coupled to hydrodynamics as discussed previously is found to be compliant with the laws of thermodynamics. We note that this prove follows a different route as the authors in \cite{cates:193} where it was assumed that different contributions to the total entropy have to be non-negative separately. Any assumption of that kind is not required in our approach.  Additionally, we allow for a general form of the entropy which is allowed to be a function of the strain variable $\bm{l}$ carrying its own entropy contribution. Furthermore, we conclude that thermodynamic consistency implies the existence of an additional entropic contribution ($\ref{GSGR9}$) to the stress tensor.

\subsection{Stress Tensor}

We can clearly distinguish the following contributions to the total pressure tensor in our GENERIC formulation of the SGR-model:

\begin{itemize}
    \item A hydrostatic pressure tensor: $\bm{\Pi}^{S}=p \bm{1}$
    \item An energetic contribution: $-k \int{\psi(E,\bm{l})\bm{l}\bm{l} }\mathrm{d}E\mathrm{d}^3\bm{l}$
    \item An entropic contribution:  \\  $\bm{\Pi}^{S}_{\text{c}}= \int{\psi(E,\bm{l})\left(2\chi\bm{1}+\chi\bm{l}\frac{\partial \ln R(E,{\bm{l}})}{\partial \bm{l}} \right) \mathrm{d}^3E }\mathrm{d}^3\bm{l}$
    \item A  Newtonian contribution from the reservoir accounting for viscous heating of the ``background'' fluid.
\end{itemize}

Note, that the entropic contribution was not considered in previous work on the SGR-model but its existence is imposed by the GENERIC structure. However, we want to stress out that this entropic contribution arises from the degeneracy condition ($\ref{GEN10}$) and a Kullback-Leibler type of ansatz for the entropy for the SGR degrees of freedom. In \cite{PhysRevE.85.031127} a different choice for this entropy was made and it might be possible that a similar entropic contribution would arise within the framework employed there, if a Kullback-Leibler entropy would be considered. \\
With our choice $R(E,\bm{l})=\rho(E,\bm{l})$ for the reference distribution, the non-isotropic part to this entropic contribution to the pressure tensor is of the following form:

\begin{equation}
\label{CON1}
        \int{\chi \psi(E,\bm{l}) \bm{l} \frac{\partial \ln \rho(E,\bm{l})}{\partial \bm{l}} \mathrm{d}^3E }\mathrm{d}^3\bm{l} \ .
\end{equation}

In tensorial models of SGR, it was suggested that $\rho(E,\bm{l})$ is proportional to $\exp(-E/T_{g})\exp(-k\bm{l}^{2}/2x)$.
With this assumption the pressure tensor contribution ($\ref{CON1}$) becomes $-k\langle \bm{ll}\rangle_{\psi}$, where we have used $x=\chi$ again.
Remarkably, this has the same form as the standard, energetic contribution to the SGR pressure tensor.
It might be a difficult task to test different (possibly similar) contributions to the stress tensor separately. If a microscopically based foundation of the SGR model going beyond its
current mean field character was available, it would be possible to test the predictions of the proposed equations numerically in simulations. Thus, tracing back the separate contributions to the stress tensor to its energetic/entropic origin remains an open problem.

\subsection{Effective Temperature $x$}
We note that the SGR model coincides with the GENERIC equations of motion if and only if we identify the effective temperature $x$ with the configurational temperature $\chi$, thus clarifying the role of this a priori undetermined parameter. Choosing $A=\chi$ in equation ($\ref{GEN9}$),  results in an equation of motion for the configurational temperature.

\begin{multline}
\label{CON2}
      \dot{\chi} = -\bm{v} \cdot \frac{\partial}{\partial \bm{r}} \chi  -  \frac{ \epsilon_{\text{c}} + 2\chi}{C_{\text{c}}^{\text{V}}} \frac{\partial}{\partial \bm{r}} \cdot \bm{v}   +  \\ \frac{1}{C_{\text{c}}^{\text{V}}} \left\{ \Big( k \langle \bm{ll} \rangle \Big) :\frac{\partial}{\partial \bm{r}}\bm{v}  +  \alpha \left(\frac{1}{\chi} -\frac{1}{T}\right) - \right. \\
\left.  \int  \phi^{E}(E,\bm{l}) \Big[- \Gamma(E,\bm{l})\psi(E,\bm{l})+  Y \rho(E,\bm{l}) \Big]  \text{d}E\text{d}^3\bm{l} \right\} \,
\end{multline}
where we introduced a configurational heat capacity at constant volume $C_{\text{c}}^{\text{V}}$. The interpretation of ($\ref{CON2}$) is as follows. The first term on the right hand side convects the configurational temperature in space. The remaining terms describe energy flows in or out of the configurational subsystem, which result in a temperature change given by the quotient of energy difference and heat capacity.
The contributions from left to right are due to work done to compress the fluid, work performed by deforming the SGR elements, heat flow between configurational subsystem and reservoir and the average energy which is dissipated as a consequence of the energy gain/loss due to the ``hopping" into another energy trap.
As a simple example we discuss the implications of ($\ref{CON2}$) for simple shear, i.e.:

\begin{equation}
\label{CON2a}
  \frac{\partial}{\partial \bm{r}} \bm{v} = \begin{pmatrix} 0 & 0 & 0 \\ \dot{\gamma}_{0} & 0 & 0 \\ 0 & 0 & 0 \\  \end{pmatrix}
\end{equation}
with the constant strain rate $\dot{\gamma}_{0}$. Considering a stationary flow, where $\dot{\psi}=0$, and assuming $\chi$ to be constant along streamlines, equation ($\ref{CON2}$) yields the following expression for the ratio $\chi/T$ as a function of $T$ and $\dot{\gamma}_{0}$:
\begin{equation}
\label{CON2b}
\frac{\chi}{T} = \frac{1}{1-2k T\dot{\gamma}_{0} \langle l_{1} l_{2} \rangle/\alpha}  
\end{equation}

Equation ($\ref{CON2b}$) means, that in this particular flow the ratio of configurational temperature $\chi$ and temperature of the reservoir $T$ is determined by the quotient of the rate at which elastic energy can be stored in the SGR-elements and the dissipation of heat into the fast subsystem.

\subsection{Entropy Production}

Employing equation ($\ref{GEN13}$), results in the following expression for the total entropy production.
\begin{multline}
\label{CON3}
   \frac{\mathrm{d}S}{\mathrm{d}t}= \int \left\{  \frac{\partial}{\partial \bm{r}} \frac{1}{T} \cdot \bm{\lambda} T^{2} \cdot \frac{\partial}{\partial \bm{r}} \frac{1}{T} + \alpha \left(\frac{1}{\chi} -\frac{1}{T}\right)^{2}    \right.  \\ \left.
  - \int  \ln \frac{\psi(E,\bm{l})}{\psi_{0}(E,\bm{l})} \Big[-\Gamma(E,\bm{l})\psi(E,\bm{l})+  Y \rho(E,\bm{l}) \Big]  \text{d}E\text{d}^3\bm{l} \right. \\ \left.
 + \left( \frac{\hat{\kappa}}{4T} \left( \mathrm{Tr} \dot{\bm{\gamma}} \right)^{2}+ \frac{\eta}{2T} \dot{\bm{\gamma}} : \dot{\bm{\gamma}} \right)  \right\} \text{d}^3\bm{r} \ ,
\end{multline}
where we have used $R(E,\bm{l})=\rho(E,\bm{l})$ again. In ($\ref{CON3}$) we can readily identify all dissipative processes contributing to the total entropy production.
We note that the form of the entropy production is considerably different from the expression given in \cite{PhysRevE.85.031127}: Firstly, the resorvoir in our formulation of SGR is not a pure heat bath but it acts as a Newtonian fluid, i.e.\ there is an entropy contribution accounting for heat transport in the reservoir, it supports stresses and can be heated via viscous heating. Secondly, besides entropy contributions arising from dissipative processes in both the reservoir and the configurational subsystem, there is an explicit term which arises from the heat exchange between the subsystems.

\subsection{SGR coupled to hydrodynamics}

The presented equations of motion provide a closed description of both thermodynamic and rheological behavior of soft glassy materials allowing for a full three dimensional treatment of hydrodynamics of these materials as it is not present in the literature (although there exists a tensorial version of SGR \cite{cates:193},  thermodynamical aspects are not treated on this general level).
The obtained equations of motion are also of local nature, slightly generalizing the original version of SGR and the supplemental equations account for a correct thermodynamic treatment.
As a conclusion we have successfully formulated the SGR model within the GENERIC framework. The obtained time evolution for the SGR distribution function $\psi$ coincides with a tensorial version of SGR. We have proven that the SGR model is thermodynamically consistent and it turned out that this consistency implies a modification to the stress tensor as stated by earlier versions of the SGR model. Additionally, our work supports the conclusion drawn in \cite{PhysRevE.85.031127} that the effective temperature $x$ as it appears in the SGR model should be identical to the configurational temperature $\chi$ associated with the slow degrees of freedom.

The present formulation of the SGR model and in particular its extension to allow for spatial inhomogeneities might be useful when studying the flow of soft glassy systems in complicated geometries. The corresponding problem of inhomogeneous deformations in hard amorphous systems has been studied only very recently via mean-field or lattice models \cite{dynhet}.
While the stress field around localized plastic events is found to have a predominant quadrupolar character, the situation is less clear-cut for soft amorphous systems.
Knowledge of the stress field and interactions between relaxation events would allow to propose an improved SGR model that goes beyond its current mean-field formulation.

\begin{acknowledgments}
  We thank Peter Sollich and Hans Christian {\"O}ttinger for insightful discussions. We also gratefully acknowledge the Swiss National Science Foundation for providing funding under Grant No.\ 200021\_134626.
\end{acknowledgments}


\appendix

\section{Construction of the Operator $M_{55}$}

In this section we discuss the construction of the operator $M_{55}$. In \cite{Schieber2003179} it was shown, that every time continous Markov process satisfying the detailed balance condition ($\ref{GSGR12}$) can be formulated within GENERIC. The corresponding operator is given by
\begin{multline}
\label{A1}
  M_{55}=\int\frac{\Delta(E,E'',\bm{l},\bm{l''})}{\sqrt{\psi_{0}\psi_{0}''}} \frac{\psi''\psi_{0}-\psi\psi_{0}''}{\ln\left( \frac{\psi''\psi_{0}}{\psi_{0}''\psi} \right)}  \times \\   \Big[\delta(E-E',\bm{l}-\bm{l'})-\delta(E''-E',\bm{l}''-\bm{l'}) \Big] \mathrm{d}E''\mathrm{d}^3\bm{l}'' \ ,
\end{multline}
where we used the fact that product of jump probability $w$ and density of states $\rho$ can be written as
\begin{equation}
\label{A2}
   w(E \rightarrow E',\bm{l} \rightarrow \bm{l}') \rho(E',\bm{l'}) = \Delta(E,E',\bm{l},\bm{l'})\sqrt{\frac{\psi_{0}}{\psi_{0}'}} \ ,
\end{equation}
with a function $\Delta$ being symmetric in both pairs of arguments ($E$,$E'$ and $\bm{l}$, $\bm{l}'$) and non-negative as  pointed out in \cite{Kampen_book}.
It is straightforward to verify that for our model this function reads as follows:
\begin{equation}
\label{A1a}
\Delta(E,E',\bm{l},\bm{l'})= \sqrt{\Gamma(E',\bm{l'}) \rho(E',\bm{l'})} \sqrt{\Gamma(E,\bm{l}) \rho(E,\bm{l})} \ ,
\end{equation}
Acting on an expression of the type $\ln \psi/\psi_{0}$, the $\delta$-distributions cancel the specific form of the logarithmic term in the denominator after an integration over the primed variables. Inserting the definition ($\ref{A2}$) in ($\ref{A1}$) yields an expression which is basically the master equation for the Markovian process.  A more detailed discussion and motivation for the specific form of $M_{55}$ can also be found in  \cite{Schieber2003179}.

\section{Symmetry and Non-negativity of $M$}
We will discuss Symmetry and Non-negativity for the three parts of $M$ separately. The first part ($M_{\text{HD}}$) is known to be positive and symmetric. \cite{PhysRevE.56.6633} The second part ($M_{\text{HF}}$) is also symmetric and its eigenvalues are $2\alpha$ and $0$. Since we assumed the heat transfer function to be positive, this part is non-negative as well. For the last part ($M_{\text{C}}$) it is sufficient to show that the operator $M_{55}$ satisfies the symmetry and positivity condition.
Firstly, we show that the operator $M_{55}$ is symmetric with respect to the transformation $\bm{l} \rightarrow \bm{l}'$, $E \rightarrow E'$.
We treat the two Kronecker-Deltas separately.

\begin{widetext}
\begin{multline}
 M_{55}=
\underbrace{\int{\frac{\Delta(E,E'',\bm{l},\bm{l''})}{\sqrt{\psi_{0}\psi_{0}''}} \frac{\psi''\psi_{0}-\psi\psi_{0}''}{\ln\left( \frac{\psi''\psi_{0}}{\psi_{0}''\psi} \right)} \delta(E-E',\bm{l}-\bm{l'})\mathrm{d}E''\mathrm{d}^3\bm{l}''}}_{a_1} - \\ \underbrace{\int{\frac{\Delta(E,E'',\bm{l},\bm{l''})}{\sqrt{\psi_{0}\psi_{0}''}} \frac{\psi''\psi_{0}-\psi\psi_{0}''}{\ln\left( \frac{\psi''\psi_{0}}{\psi_{0}''\psi} \right)} \delta(E''-E',\bm{l}''-\bm{l'})  \mathrm{d}E''\mathrm{d}^3\bm{l}''}}_{a_2}
\end{multline}
\end{widetext}

The whole first expression inherits the symmetry of the Kronecker-Delta. Performing the integration in the second part yields:

\begin{equation}
\label{GSGR15}
 a_{2} \sim \frac{\Delta(E,E',\bm{l},\bm{l'})}{\sqrt{\psi_{0}\psi_{0}'}} \frac{\psi'\psi_{0}-\psi\psi_{0}'}{\ln\left( \frac{\psi'\psi_{0}}{\psi_{0}'\psi} \right)}
\end{equation}
The first factor is obviously symmetric. The second factor is a quotient of two antisymmetric expressions and therefore symmetric as well. \\

In order to prove the positivity of $M_{55}$, we rewrite it in the following form:
\begin{widetext}
\begin{multline}
\label{GSGR16}
 M_{55}= \frac{1}{2} \int\int\overbrace{\frac{\Delta(F,F',\bm{k},\bm{k'})}{\sqrt{\psi_{0}(F,\bm{k})\psi_{0}(F',\bm{k'})}}}^{b_1}  \overbrace{\frac{\psi(F',\bm{k'})\psi_{0}(F,\bm{k})-\psi(F,\bm{k})\psi_{0}(F',\bm{k'})}{\ln\left( \frac{\psi(F',\bm{k'})\psi_{0}(F,\bm{k})}{\psi_{0}(F',\bm{k'})\psi(F,\bm{k})} \right)}}^{b_2}    \begin{pmatrix} \delta(F-E,\bm{k}-\bm{l}) \\ \delta(F'-E,\bm{k}'-\bm{l}) \end{pmatrix}   \cdot  \\ \underbrace{\begin{pmatrix} 1 & -1 \\ -1 & 1 \end{pmatrix}}_{b_3} \cdot \begin{pmatrix} \delta(F-E',\bm{k}-\bm{l}') \\ \delta(F'-E',\bm{k}'-\bm{l}') \end{pmatrix} \mathrm{d}F\mathrm{d}^3\bm{k}\mathrm{d}F'\mathrm{d}^3\bm{k}'
\end{multline}
\end{widetext}
Using the symmetry properties discussed previously, it is straightforward to verify that this is indeed the operator $M_{55}$.
The expressions $b_1$ and $b_2$ are necessarily positive and the matrix $b_3$ has the eigenvalues $2$ (eigenvector $\left(-1,1\right)$) and $0$ (eigenvector $\left(1,1\right)$).
Therefore the operator $M_{55}$ is nonnegative.

\bibliography{if_SGR_ref}

\begin{thebibliography}{26}%
\makeatletter
\providecommand \@ifxundefined [1]{%
 \@ifx{#1\undefined}
}%
\providecommand \@ifnum [1]{%
 \ifnum #1\expandafter \@firstoftwo
 \else \expandafter \@secondoftwo
 \fi
}%
\providecommand \@ifx [1]{%
 \ifx #1\expandafter \@firstoftwo
 \else \expandafter \@secondoftwo
 \fi
}%
\providecommand \natexlab [1]{#1}%
\providecommand \enquote  [1]{``#1''}%
\providecommand \bibnamefont  [1]{#1}%
\providecommand \bibfnamefont [1]{#1}%
\providecommand \citenamefont [1]{#1}%
\providecommand \href@noop [0]{\@secondoftwo}%
\providecommand \href [0]{\begingroup \@sanitize@url \@href}%
\providecommand \@href[1]{\@@startlink{#1}\@@href}%
\providecommand \@@href[1]{\endgroup#1\@@endlink}%
\providecommand \@sanitize@url [0]{\catcode `\\12\catcode `\$12\catcode
  `\&12\catcode `\#12\catcode `\^12\catcode `\_12\catcode `\%12\relax}%
\providecommand \@@startlink[1]{}%
\providecommand \@@endlink[0]{}%
\providecommand \url  [0]{\begingroup\@sanitize@url \@url }%
\providecommand \@url [1]{\endgroup\@href {#1}{\urlprefix }}%
\providecommand \urlprefix  [0]{URL }%
\providecommand \Eprint [0]{\href }%
\providecommand \doibase [0]{http://dx.doi.org/}%
\providecommand \selectlanguage [0]{\@gobble}%
\providecommand \bibinfo  [0]{\@secondoftwo}%
\providecommand \bibfield  [0]{\@secondoftwo}%
\providecommand \translation [1]{[#1]}%
\providecommand \BibitemOpen [0]{}%
\providecommand \bibitemStop [0]{}%
\providecommand \bibitemNoStop [0]{.\EOS\space}%
\providecommand \EOS [0]{\spacefactor3000\relax}%
\providecommand \BibitemShut  [1]{\csname bibitem#1\endcsname}%
\let\auto@bib@innerbib\@empty
\bibitem [{\citenamefont {H\'ebraud}\ and\ \citenamefont
  {Lequeux}(1998)}]{PhysRevLett.81.2934}%
  \BibitemOpen
  \bibfield  {author} {\bibinfo {author} {\bibfnamefont {P.}~\bibnamefont
  {H\'ebraud}}\ and\ \bibinfo {author} {\bibfnamefont {F.}~\bibnamefont
  {Lequeux}},\ }\href {\doibase 10.1103/PhysRevLett.81.2934} {\bibfield
  {journal} {\bibinfo  {journal} {Phys. Rev. Lett.}\ }\textbf {\bibinfo
  {volume} {81}},\ \bibinfo {pages} {2934} (\bibinfo {year}
  {1998})}\BibitemShut {NoStop}%
\bibitem [{\citenamefont {Picard}\ \emph {et~al.}(2002)\citenamefont {Picard},
  \citenamefont {Ajdari}, \citenamefont {Bocquet},\ and\ \citenamefont
  {Lequeux}}]{PhysRevE.66.051501}%
  \BibitemOpen
  \bibfield  {author} {\bibinfo {author} {\bibfnamefont {G.}~\bibnamefont
  {Picard}}, \bibinfo {author} {\bibfnamefont {A.}~\bibnamefont {Ajdari}},
  \bibinfo {author} {\bibfnamefont {L.}~\bibnamefont {Bocquet}}, \ and\
  \bibinfo {author} {\bibfnamefont {F.}~\bibnamefont {Lequeux}},\ }\href
  {\doibase 10.1103/PhysRevE.66.051501} {\bibfield  {journal} {\bibinfo
  {journal} {Phys. Rev. E}\ }\textbf {\bibinfo {volume} {66}},\ \bibinfo
  {pages} {051501} (\bibinfo {year} {2002})}\BibitemShut {NoStop}%
\bibitem [{\citenamefont {Bocquet}\ \emph {et~al.}(2009)\citenamefont
  {Bocquet}, \citenamefont {Colin},\ and\ \citenamefont
  {Ajdari}}]{PhysRevLett.103.036001}%
  \BibitemOpen
  \bibfield  {author} {\bibinfo {author} {\bibfnamefont {L.}~\bibnamefont
  {Bocquet}}, \bibinfo {author} {\bibfnamefont {A.}~\bibnamefont {Colin}}, \
  and\ \bibinfo {author} {\bibfnamefont {A.}~\bibnamefont {Ajdari}},\ }\href
  {\doibase 10.1103/PhysRevLett.103.036001} {\bibfield  {journal} {\bibinfo
  {journal} {Phys. Rev. Lett.}\ }\textbf {\bibinfo {volume} {103}},\ \bibinfo
  {pages} {036001} (\bibinfo {year} {2009})}\BibitemShut {NoStop}%
\bibitem [{\citenamefont {Mansard}\ \emph {et~al.}(2011)\citenamefont
  {Mansard}, \citenamefont {Colin}, \citenamefont {Chauduri},\ and\
  \citenamefont {Bocquet}}]{C1SM05229B}%
  \BibitemOpen
  \bibfield  {author} {\bibinfo {author} {\bibfnamefont {V.}~\bibnamefont
  {Mansard}}, \bibinfo {author} {\bibfnamefont {A.}~\bibnamefont {Colin}},
  \bibinfo {author} {\bibfnamefont {P.}~\bibnamefont {Chauduri}}, \ and\
  \bibinfo {author} {\bibfnamefont {L.}~\bibnamefont {Bocquet}},\ }\href
  {\doibase 10.1039/C1SM05229B} {\bibfield  {journal} {\bibinfo  {journal}
  {Soft Matter}\ }\textbf {\bibinfo {volume} {7}},\ \bibinfo {pages} {5524}
  (\bibinfo {year} {2011})}\BibitemShut {NoStop}%
\bibitem [{\citenamefont {Talamali}\ \emph {et~al.}(2008)\citenamefont
  {Talamali}, \citenamefont {Pet\"aj\"a}, \citenamefont {Vandembroucq},\ and\
  \citenamefont {Roux}}]{PhysRevE.78.016109}%
  \BibitemOpen
  \bibfield  {author} {\bibinfo {author} {\bibfnamefont {M.}~\bibnamefont
  {Talamali}}, \bibinfo {author} {\bibfnamefont {V.}~\bibnamefont
  {Pet\"aj\"a}}, \bibinfo {author} {\bibfnamefont {D.}~\bibnamefont
  {Vandembroucq}}, \ and\ \bibinfo {author} {\bibfnamefont {S.}~\bibnamefont
  {Roux}},\ }\href {\doibase 10.1103/PhysRevE.78.016109} {\bibfield  {journal}
  {\bibinfo  {journal} {Phys. Rev. E}\ }\textbf {\bibinfo {volume} {78}},\
  \bibinfo {pages} {016109} (\bibinfo {year} {2008})}\BibitemShut {NoStop}%
\bibitem [{\citenamefont {Pouliquen}\ and\ \citenamefont
  {Forterre}(2009)}]{Pouliquen28122009}%
  \BibitemOpen
  \bibfield  {author} {\bibinfo {author} {\bibfnamefont {O.}~\bibnamefont
  {Pouliquen}}\ and\ \bibinfo {author} {\bibfnamefont {Y.}~\bibnamefont
  {Forterre}},\ }\href@noop {} {\bibfield  {journal} {\bibinfo  {journal}
  {Philosophical Transactions of the Royal Society A: Mathematical, Physical
  and Engineering Sciences}\ }\textbf {\bibinfo {volume} {367}},\ \bibinfo
  {pages} {5091} (\bibinfo {year} {2009})}\BibitemShut {NoStop}%
\bibitem [{\citenamefont {Sollich}\ \emph {et~al.}(1997)\citenamefont
  {Sollich}, \citenamefont {Lequeux}, \citenamefont {H\'ebraud},\ and\
  \citenamefont {Cates}}]{SGR}%
  \BibitemOpen
  \bibfield  {author} {\bibinfo {author} {\bibfnamefont {P.}~\bibnamefont
  {Sollich}}, \bibinfo {author} {\bibfnamefont {F.}~\bibnamefont {Lequeux}},
  \bibinfo {author} {\bibfnamefont {P.}~\bibnamefont {H\'ebraud}}, \ and\
  \bibinfo {author} {\bibfnamefont {M.~E.}\ \bibnamefont {Cates}},\ }\href@noop
  {} {\bibfield  {journal} {\bibinfo  {journal} {Phys. Rev. Lett.}\ }\textbf
  {\bibinfo {volume} {78}},\ \bibinfo {pages} {2020} (\bibinfo {year}
  {1997})}\BibitemShut {NoStop}%
\bibitem [{\citenamefont {Orsi}\ \emph {et~al.}(2012)\citenamefont {Orsi},
  \citenamefont {Baldi}, \citenamefont {Cicuta},\ and\ \citenamefont
  {Cristofolini}}]{Orsi201271}%
  \BibitemOpen
  \bibfield  {author} {\bibinfo {author} {\bibfnamefont {D.}~\bibnamefont
  {Orsi}}, \bibinfo {author} {\bibfnamefont {G.}~\bibnamefont {Baldi}},
  \bibinfo {author} {\bibfnamefont {P.}~\bibnamefont {Cicuta}}, \ and\ \bibinfo
  {author} {\bibfnamefont {L.}~\bibnamefont {Cristofolini}},\ }\href@noop {}
  {\bibfield  {journal} {\bibinfo  {journal} {Colloids and Surfaces A:
  Physicochemical and Engineering Aspects}\ }\textbf {\bibinfo {volume}
  {413}},\ \bibinfo {pages} {71 } (\bibinfo {year} {2012})}\BibitemShut
  {NoStop}%
\bibitem [{\citenamefont {Srivastava}\ \emph {et~al.}(2012)\citenamefont
  {Srivastava}, \citenamefont {Shin},\ and\ \citenamefont
  {Archer}}]{C2SM06889C}%
  \BibitemOpen
  \bibfield  {author} {\bibinfo {author} {\bibfnamefont {S.}~\bibnamefont
  {Srivastava}}, \bibinfo {author} {\bibfnamefont {J.~H.}\ \bibnamefont
  {Shin}}, \ and\ \bibinfo {author} {\bibfnamefont {L.~A.}\ \bibnamefont
  {Archer}},\ }\href {\doibase 10.1039/C2SM06889C} {\bibfield  {journal}
  {\bibinfo  {journal} {Soft Matter}\ }\textbf {\bibinfo {volume} {8}},\
  \bibinfo {pages} {4097} (\bibinfo {year} {2012})}\BibitemShut {NoStop}%
\bibitem [{\citenamefont {Agarwal}\ \emph {et~al.}(2011)\citenamefont
  {Agarwal}, \citenamefont {Srivastava},\ and\ \citenamefont
  {Archer}}]{PhysRevLett.107.268302}%
  \BibitemOpen
  \bibfield  {author} {\bibinfo {author} {\bibfnamefont {P.}~\bibnamefont
  {Agarwal}}, \bibinfo {author} {\bibfnamefont {S.}~\bibnamefont {Srivastava}},
  \ and\ \bibinfo {author} {\bibfnamefont {L.~A.}\ \bibnamefont {Archer}},\
  }\href {\doibase 10.1103/PhysRevLett.107.268302} {\bibfield  {journal}
  {\bibinfo  {journal} {Phys. Rev. Lett.}\ }\textbf {\bibinfo {volume} {107}},\
  \bibinfo {pages} {268302} (\bibinfo {year} {2011})}\BibitemShut {NoStop}%
\bibitem [{\citenamefont {Srivastava}\ \emph {et~al.}(2011)\citenamefont
  {Srivastava}, \citenamefont {Leiske}, \citenamefont {Basu},\ and\
  \citenamefont {Fuller}}]{C0SM00839G}%
  \BibitemOpen
  \bibfield  {author} {\bibinfo {author} {\bibfnamefont {S.}~\bibnamefont
  {Srivastava}}, \bibinfo {author} {\bibfnamefont {D.}~\bibnamefont {Leiske}},
  \bibinfo {author} {\bibfnamefont {J.~K.}\ \bibnamefont {Basu}}, \ and\
  \bibinfo {author} {\bibfnamefont {G.~G.}\ \bibnamefont {Fuller}},\ }\href
  {\doibase 10.1039/C0SM00839G} {\bibfield  {journal} {\bibinfo  {journal}
  {Soft Matter}\ }\textbf {\bibinfo {volume} {7}},\ \bibinfo {pages} {1994}
  (\bibinfo {year} {2011})}\BibitemShut {NoStop}%
\bibitem [{\citenamefont {Evans}\ \emph {et~al.}(1999)\citenamefont {Evans},
  \citenamefont {Cates},\ and\ \citenamefont {Sollich}}]{DiffSGR}%
  \BibitemOpen
  \bibfield  {author} {\bibinfo {author} {\bibfnamefont {R.}~\bibnamefont
  {Evans}}, \bibinfo {author} {\bibfnamefont {M.}~\bibnamefont {Cates}}, \ and\
  \bibinfo {author} {\bibfnamefont {P.}~\bibnamefont {Sollich}},\ }\href
  {\doibase 10.1007/s100510050902} {\bibfield  {journal} {\bibinfo  {journal}
  {The European Physical Journal B - Condensed Matter and Complex Systems}\
  }\textbf {\bibinfo {volume} {10}},\ \bibinfo {pages} {705} (\bibinfo {year}
  {1999})}\BibitemShut {NoStop}%
\bibitem [{\citenamefont {Sollich}(1998)}]{PhysRevE.58.738}%
  \BibitemOpen
  \bibfield  {author} {\bibinfo {author} {\bibfnamefont {P.}~\bibnamefont
  {Sollich}},\ }\href {\doibase 10.1103/PhysRevE.58.738} {\bibfield  {journal}
  {\bibinfo  {journal} {Phys. Rev. E}\ }\textbf {\bibinfo {volume} {58}},\
  \bibinfo {pages} {738} (\bibinfo {year} {1998})}\BibitemShut {NoStop}%
\bibitem [{\citenamefont {Fielding}\ \emph {et~al.}(2000)\citenamefont
  {Fielding}, \citenamefont {Sollich},\ and\ \citenamefont
  {Cates}}]{fielding:323}%
  \BibitemOpen
  \bibfield  {author} {\bibinfo {author} {\bibfnamefont {S.~M.}\ \bibnamefont
  {Fielding}}, \bibinfo {author} {\bibfnamefont {P.}~\bibnamefont {Sollich}}, \
  and\ \bibinfo {author} {\bibfnamefont {M.~E.}\ \bibnamefont {Cates}},\ }\href
  {\doibase 10.1122/1.551088} {\bibfield  {journal} {\bibinfo  {journal}
  {Journal of Rheology}\ }\textbf {\bibinfo {volume} {44}},\ \bibinfo {pages}
  {323} (\bibinfo {year} {2000})}\BibitemShut {NoStop}%
\bibitem [{\citenamefont {Sollich}\ and\ \citenamefont
  {Cates}(2012)}]{PhysRevE.85.031127}%
  \BibitemOpen
  \bibfield  {author} {\bibinfo {author} {\bibfnamefont {P.}~\bibnamefont
  {Sollich}}\ and\ \bibinfo {author} {\bibfnamefont {M.~E.}\ \bibnamefont
  {Cates}},\ }\href {\doibase 10.1103/PhysRevE.85.031127} {\bibfield  {journal}
  {\bibinfo  {journal} {Phys. Rev. E}\ }\textbf {\bibinfo {volume} {85}},\
  \bibinfo {pages} {031127} (\bibinfo {year} {2012})}\BibitemShut {NoStop}%
\bibitem [{\citenamefont {Bouchbinder}\ and\ \citenamefont
  {Langer}(2009{\natexlab{a}})}]{PhysRevE.80.031131}%
  \BibitemOpen
  \bibfield  {author} {\bibinfo {author} {\bibfnamefont {E.}~\bibnamefont
  {Bouchbinder}}\ and\ \bibinfo {author} {\bibfnamefont {J.~S.}\ \bibnamefont
  {Langer}},\ }\href {\doibase 10.1103/PhysRevE.80.031131} {\bibfield
  {journal} {\bibinfo  {journal} {Phys. Rev. E}\ }\textbf {\bibinfo {volume}
  {80}},\ \bibinfo {pages} {031131} (\bibinfo {year}
  {2009}{\natexlab{a}})}\BibitemShut {NoStop}%
\bibitem [{\citenamefont {Bouchbinder}\ and\ \citenamefont
  {Langer}(2009{\natexlab{b}})}]{PhysRevE.80.031132}%
  \BibitemOpen
  \bibfield  {author} {\bibinfo {author} {\bibfnamefont {E.}~\bibnamefont
  {Bouchbinder}}\ and\ \bibinfo {author} {\bibfnamefont {J.~S.}\ \bibnamefont
  {Langer}},\ }\href {\doibase 10.1103/PhysRevE.80.031132} {\bibfield
  {journal} {\bibinfo  {journal} {Phys. Rev. E}\ }\textbf {\bibinfo {volume}
  {80}},\ \bibinfo {pages} {031132} (\bibinfo {year}
  {2009}{\natexlab{b}})}\BibitemShut {NoStop}%
\bibitem [{\citenamefont {Bouchbinder}\ and\ \citenamefont
  {Langer}(2009{\natexlab{c}})}]{PhysRevE.80.031133}%
  \BibitemOpen
  \bibfield  {author} {\bibinfo {author} {\bibfnamefont {E.}~\bibnamefont
  {Bouchbinder}}\ and\ \bibinfo {author} {\bibfnamefont {J.~S.}\ \bibnamefont
  {Langer}},\ }\href {\doibase 10.1103/PhysRevE.80.031133} {\bibfield
  {journal} {\bibinfo  {journal} {Phys. Rev. E}\ }\textbf {\bibinfo {volume}
  {80}},\ \bibinfo {pages} {031133} (\bibinfo {year}
  {2009}{\natexlab{c}})}\BibitemShut {NoStop}%
\bibitem [{\citenamefont {\"Ottinger}(2005)}]{hco_book}%
  \BibitemOpen
  \bibfield  {author} {\bibinfo {author} {\bibfnamefont {H.~C.}\ \bibnamefont
  {\"Ottinger}},\ }\href@noop {} {\emph {\bibinfo {title} {Beyond Equilibrium
  Thermodynamics}}}\ (\bibinfo  {publisher} {Wiley},\ \bibinfo {year}
  {2005})\BibitemShut {NoStop}%
\bibitem [{\citenamefont {Cates}\ and\ \citenamefont
  {Sollich}(2004)}]{cates:193}%
  \BibitemOpen
  \bibfield  {author} {\bibinfo {author} {\bibfnamefont {M.~E.}\ \bibnamefont
  {Cates}}\ and\ \bibinfo {author} {\bibfnamefont {P.}~\bibnamefont
  {Sollich}},\ }\href {\doibase 10.1122/1.1634985} {\bibfield  {journal}
  {\bibinfo  {journal} {Journal of Rheology}\ }\textbf {\bibinfo {volume}
  {48}},\ \bibinfo {pages} {193} (\bibinfo {year} {2004})}\BibitemShut
  {NoStop}%
\bibitem [{\citenamefont {Grmela}\ and\ \citenamefont
  {\"Ottinger}(1997)}]{PhysRevE.56.6620}%
  \BibitemOpen
  \bibfield  {author} {\bibinfo {author} {\bibfnamefont {M.}~\bibnamefont
  {Grmela}}\ and\ \bibinfo {author} {\bibfnamefont {H.~C.}\ \bibnamefont
  {\"Ottinger}},\ }\href {\doibase 10.1103/PhysRevE.56.6620} {\bibfield
  {journal} {\bibinfo  {journal} {Phys. Rev. E}\ }\textbf {\bibinfo {volume}
  {56}},\ \bibinfo {pages} {6620} (\bibinfo {year} {1997})}\BibitemShut
  {NoStop}%
\bibitem [{\citenamefont {\"Ottinger}\ and\ \citenamefont
  {Grmela}(1997)}]{PhysRevE.56.6633}%
  \BibitemOpen
  \bibfield  {author} {\bibinfo {author} {\bibfnamefont {H.~C.}\ \bibnamefont
  {\"Ottinger}}\ and\ \bibinfo {author} {\bibfnamefont {M.}~\bibnamefont
  {Grmela}},\ }\href {\doibase 10.1103/PhysRevE.56.6633} {\bibfield  {journal}
  {\bibinfo  {journal} {Phys. Rev. E}\ }\textbf {\bibinfo {volume} {56}},\
  \bibinfo {pages} {6633} (\bibinfo {year} {1997})}\BibitemShut {NoStop}%
\bibitem [{\citenamefont {H\"utter}\ \emph {et~al.}(2009)\citenamefont
  {H\"utter}, \citenamefont {Grmela},\ and\ \citenamefont
  {\"Ottinger}}]{hcogrmelahutter}%
  \BibitemOpen
  \bibfield  {author} {\bibinfo {author} {\bibfnamefont {M.}~\bibnamefont
  {H\"utter}}, \bibinfo {author} {\bibfnamefont {M.}~\bibnamefont {Grmela}}, \
  and\ \bibinfo {author} {\bibfnamefont {H.}~\bibnamefont {\"Ottinger}},\
  }\href {\doibase 10.1007/s00397-009-0371-y} {\bibfield  {journal} {\bibinfo
  {journal} {Rheologica Acta}\ }\textbf {\bibinfo {volume} {48}},\ \bibinfo
  {pages} {769} (\bibinfo {year} {2009})}\BibitemShut {NoStop}%
\bibitem [{\citenamefont {Schieber}(2003)}]{Schieber2003179}%
  \BibitemOpen
  \bibfield  {author} {\bibinfo {author} {\bibfnamefont {J.}~\bibnamefont
  {Schieber}},\ }\href@noop {} {\bibfield  {journal} {\bibinfo  {journal}
  {Journal of Non-Equilibrium Thermodynamics}\ }\textbf {\bibinfo {volume}
  {28}},\ \bibinfo {pages} {179} (\bibinfo {year} {2003})}\BibitemShut
  {NoStop}%
\bibitem [{\citenamefont {Barrat}\ and\ \citenamefont
  {Lemaitre}(2011)}]{dynhet}%
  \BibitemOpen
  \bibfield  {author} {\bibinfo {author} {\bibfnamefont {J.-L.}\ \bibnamefont
  {Barrat}}\ and\ \bibinfo {author} {\bibfnamefont {A.}~\bibnamefont
  {Lemaitre}},\ }in\ \href@noop {} {\emph {\bibinfo {booktitle} {Dynamical
  Heterogeneities in Glasses, Colloids, and Granular Media}}},\ \bibinfo
  {editor} {edited by\ \bibinfo {editor} {\bibfnamefont {L.}~\bibnamefont
  {Berthier}}, \bibinfo {editor} {\bibfnamefont {G.}~\bibnamefont {Biroli}},
  \bibinfo {editor} {\bibfnamefont {J.-P.}\ \bibnamefont {Bouchaud}}, \bibinfo
  {editor} {\bibfnamefont {L.}~\bibnamefont {Cipelletti}}, \ and\ \bibinfo
  {editor} {\bibfnamefont {W.}~\bibnamefont {van Saarloos}}}\ (\bibinfo
  {publisher} {Oxford University Press},\ \bibinfo {year} {2011})\BibitemShut
  {NoStop}%
\bibitem [{\citenamefont {van Kampen}(1992)}]{Kampen_book}%
  \BibitemOpen
  \bibfield  {author} {\bibinfo {author} {\bibfnamefont {N.}~\bibnamefont {van
  Kampen}},\ }\href@noop {} {\emph {\bibinfo {title} {Stochastic Processes in
  Physics and Chemistry}}}\ (\bibinfo  {publisher} {Elsevier},\ \bibinfo {year}
  {1992})\BibitemShut {NoStop}%
\end{thebibliography}%

\end{document}